# THE FUTURE OF DATA SCIENCE EDUCATION

Loreto Peter Alonzi III, School of Data Science, University of Virginia, 400 Brandon Ave, Charlottesville, VA 22903, 434-924-7835, alonzi@virginia.edu
Brian Wright, School of Data Science, University of Virginia, 400 Brandon Ave, Charlottesville, VA 22903, 434-982-2600, brianwright@virginia.edu
Ali Rivera, School of Data Science, University of Virginia, 400 Brandon Ave, Charlottesville, VA 22903, 434-982-2600

## ABSTRACT

The definition of “Data Science“ is a hotly debated topic. For many, the definition is a simple shortcut to Artificial Intelligence or Machine Learning. However there is far more depth and nuance to the field of Data Science than a simple shortcut can provide. The School of Data Science at the University of Virginia has developed a novel model for the definition of Data Science. This model is based on identifying a unified understanding of the data work done across all areas of Data Science. It represents a generational leap forward in how we understand and teach Data Science. In this paper we will present the core features of the model and explain how it unifies various concepts going far beyond the analytics component of AI. From this foundation we will present our Undergraduate Major curriculum in Data Science and demonstrate how it prepares students to be well-rounded Data Science team members and leaders. The paper will conclude with an in depth overview of the Foundations of Data Science course designed to introduce students to the field while also implementing proven STEM oriented pedagogical methods. These include, for example, specifications grading, active learning lectures, guest lectures from industry experts and weekly gamification labs.

## BACKGROUND TO THE FIELD OF DATA SCIENCE

### Introduction

The field of Data Science has thrown itself onto the landscape of academia at what seems like an unprecedented pace. The consequences of this accelerated immersion is a wide variety of organizational structures supporting various forms of Data Science initiatives. This includes everything from single departments inside pre-existing colleges/schools to stand-alone schools with dedicated Deans and tenurable faculty.  These varying approaches have produced significantly different academic programs and curriculum, with seemingly no two schools supporting identical coursework. This is nowhere more evident than with a simple review of introductory data science courses from across the US higher education landscape. The results of which show a variety of topics and requirements that seem to be able to span several disciplines including Statistics and Computer Science. It is a result of this wide spectrum that the School of Data Science at the University of Virginia worked to establish a unifying vision of what a Data Science curriculum should encompass, free of prior influence. Rafael Alvardo outlines this perspective in his essay “The 4+1 Model of Data Science” [1].  Once this vision was finalized and agreed upon, the high level structure became the foundation to the development of academic programs to include the creation of a PhD and Bachelor of Science degree. The following will provide a bit of history of the field of Data Science, map out the Virginia Model of Data Science (4+1), detail the Bachelor of Science in Data Science, and end with what we believe is a natural evolution of introductory data science courses, to include both a more narrowed and expanded view of the field that better represents how Data Science is actually practiced outside the walls of academia.

**Brief History of the Field**

The term Data Science has actually been around for more than 60 years. The two word combination was first seen in 1962 with the creation of a Data Science Laboratory by the US Air Force in Cambridge. The focus of the (DSL) was to organize the effort that had been generated since the end of WWII in advances in computation, data collection and usage. From this initial seed a series of government and academic collaborations would simmer the field forward for the next 40 years. This includes the creation of the International Council for Science Committee (ICSU) on Data for Science and Technology (CODATA) in 1966 which is still present today and founder of the Data Science Journal in 2001[2].

These several decades worth of academic simmer often came in the representation of tensions between the field of Statistics and more algorithmically driven approaches in Computer Science. A rather well known representation of this tension came from Leo Briemen in his 2001 essay "Two Cultures" [3]. Briemen suggests that the field of statistics works to collect data and model real world scenarios whereas the data mining or algorithmic driven approaches uses data first without consideration of underlying realities. While Briemen was a Statistician and had his biases, the suggestion of two rapidly evolving cultures was correct. Traditional approaches would design data gathering around a problem foregrounding the hypothesis at hand whereas data mining (data science) approaches often go searching for a problem that can be answered with data already in hand.

This separation represents a turning point in the field where data analytics as an antecedent to the formal field of Data Science really starts to take hold, largely driven by private sector interests. It is during this period of development that a significant increase in the call for expanding educational offerings to include data skills more generally, in addition to commonly offered inferential or econometric methods was heard [2]. This 50 year journey came to head in 2008 with the publication of a Harvard Business Review article entitled: "Data Scientist: The Sexiest Job of the 21st Century." The article detailed the use of the term in Silicon Valley startups, in particular Facebook and Linkedin, for what their employees were doing on the ground. The article also suggested, not so subtly, that the field was ripe for explosion [4].

The prediction could not have been more accurate. This moment represents a point in time where private sector demand and the long running academic evolution seemed to meet for a uniquely beneficial moment. The result is an unprecedented growth of a new academic field.

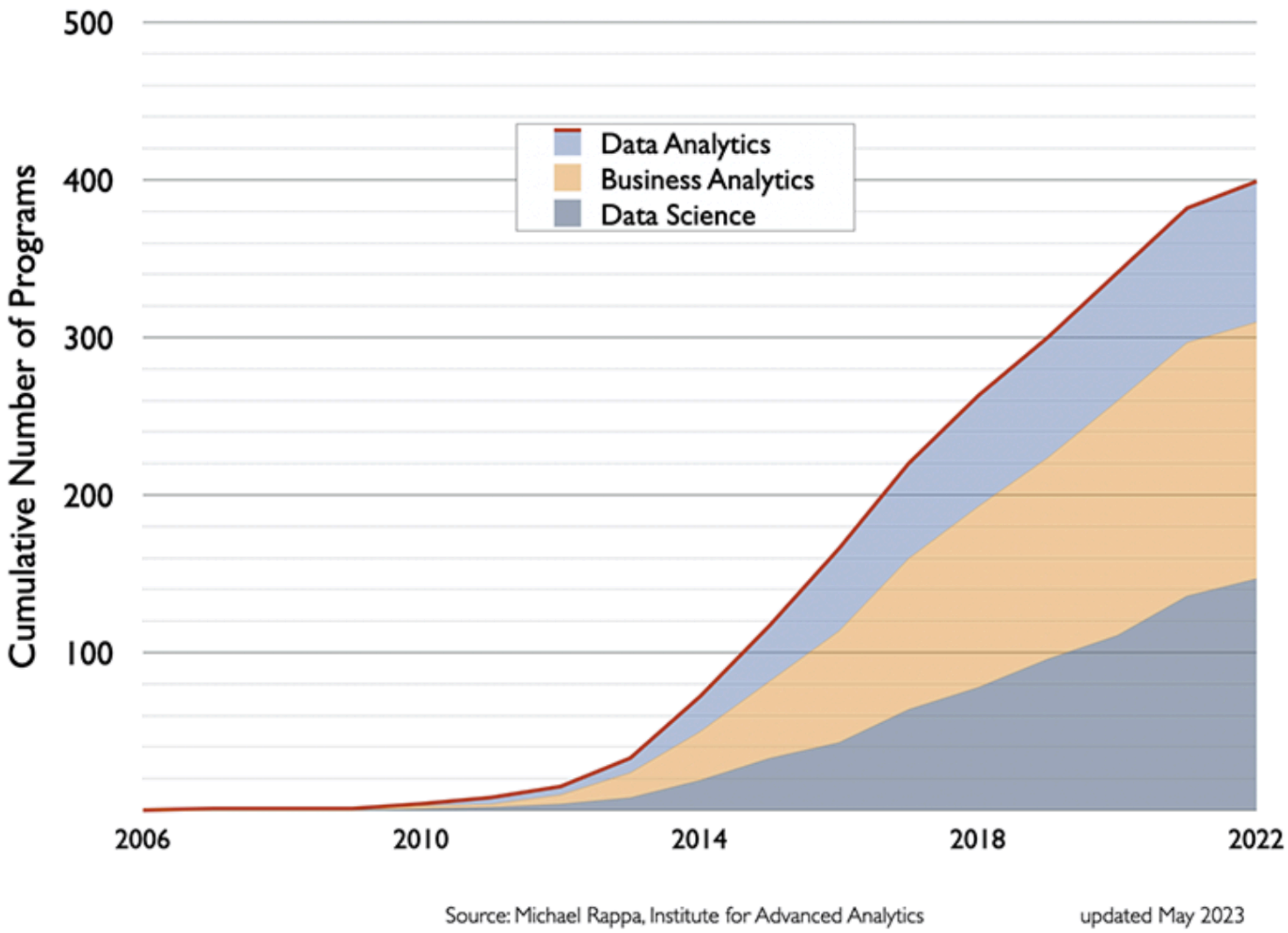

**FIGURE 1: The Growth of Analytics and Data Science Master's Degree Programs in the United States (2006-2022)**

Michael Rappa, the Director of the Advanced Analytics Center at NC State, was one of the first programs in the country to formally introduce what we know of today as a Data Science curriculum at the graduate level, doing so in the early 2000s. Consequently, he has had a unique perspective on the growth of the field and has tracked the expansion of graduate programs over the last 15 years. Figure 1 shows the explosion of graduate programs between 2010 and 2022 [5]. Driven by this collision of private sector demand and the academic maturity of Data Science. This demand for growth, however, came at a cost. The field has been rolled out onto campuses at such a rapid pace that it seems no two programs have significant overlap, both as it relates to organizational structure and curricular content. Nowhere is this more clear than a review of Introductory Data Science courses.

Figure 2 shows text analysis conducted by our team on sampled syllabi for 40 current Introduction to Data Science courses from four year institutions across the US (collected in 2023) [6]. We worked to limit the inclusion of courses to only those with "Data" and "Science" in the title or courses that were known to have a Data Science orientation. Berkeley's Data 8, for example, is included despite having no "Science" in the title. Fuzzy clusters were extracted from the corpus by using Latent Dirichlet Allocation (LDA). These clusters were then visualized using word clouds with the beta values dictating size, as opposed to the traditional use of word count. The output was well separated into four categories.

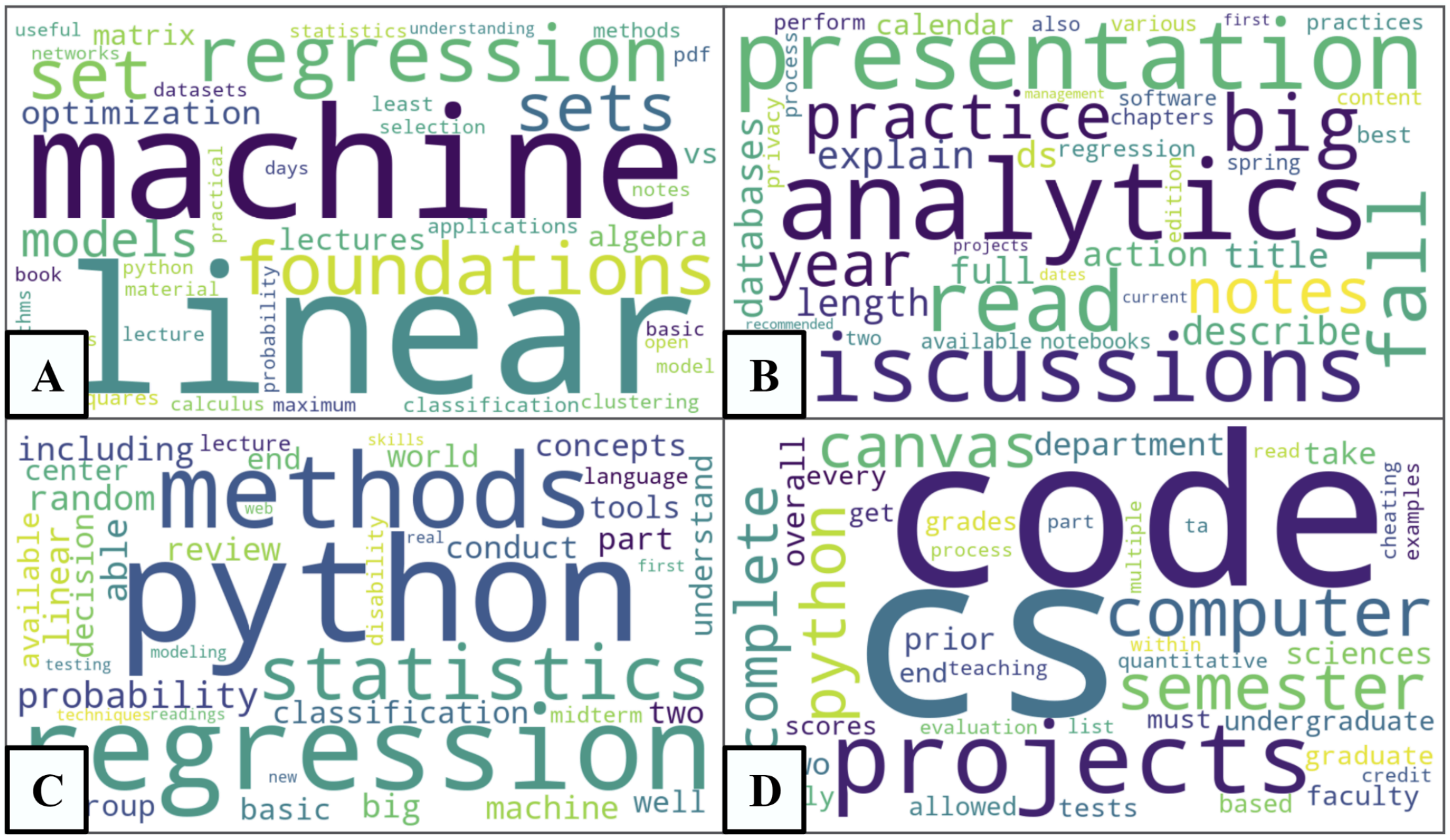

**FIGURE 2: Fuzzy clusters derived from 40 syllabi using Latent Dirichlet Allocation (LDA).**

Figure 2-A represents courses focused exclusively on analytical approaches, likely machine learning starting with linear regression. Figure 2-B emphasizes data presentations and communications while focusing still on analytics primarily. The third cluster, Figure 2- C has much more of a focus on programming with python specifically, in combination with traditional statistical methods likely offered through a statistics or mathematics department. The final cluster, Figure 2-D seems exclusively focused on programming/coding seemingly offered through a Computer Science department.  This analysis, while limited, gives us a nice initial sense of the variety of approaches used for introductory courses.  We can infer that these courses inherent content as a by-product of the department offering the courses, with CS courses focusing more on programming and Mathematics/Statistics focusing more on introductory analytical methods like linear regression. However, more importantly than what is present is what is not.  In none of the courses is there any significant mention of data ethics or databases or data visualization. This is not overly surprising given how the field exploded onto college campuses. It is almost certainly the case that the pressure to begin Data Science programs meant that universities took advantage of current resources whenever possible. Which meant repackaging current courses or content to have a bit more Data Science flavor while still being heavily influenced from the original content.

The results of this review, though not surprising, does raise some questions about how the field is being represented. While it is hard to argue against the core being analytical approaches delivered through programming, focusing on those topics exclusively is reductive to what Data Science has been growing towards as a unique and independent discipline. As an example, there is a legitimate argument to be made that knowing and understanding data systems (structures, database, querying languages, etc.) and/or the complicated ethical issues of the field should come before entering analytics all together. Regardless of the order with which topics are presented this representation of the field appears incomplete.

The School of Data Science at the University of Virginia was the first stand-alone School of its kind in the US. As a result, we had the luxury (and burden) of designing our Data Science curriculum from the ground up. In doing so, Rafael Alvardo, a Professor in the School, worked in collaboration with the first wave of faculty to create a framework for the field. This framework was the DNA utilized over the last three years to generate the PhD and undergraduate programs. The following section will provide an overview of the framework followed by details on how it influenced the development of the undergraduate program and specifically the Foundations of Data Science course.

**University of Virginia Model of Data Science**

UVA's areas of Data Science model (4+1) was generated by Rafael Alvarado in collaboration with Data Science faculty in 2019 leading up to the official beginning of the School of Data Science in the fall of 2019. The framework is designed to represent areas of focus or areas that should be foundation to all data science curriculum. These include Value, Design, Systems, and Analytics. These areas are umbrella concepts that could and should encompass a variety of more specific learning outcomes dependent on context. However, implicit in the structure is that an effort to balance the curricular content across the four areas is critical, as each contains necessary aspects for becoming a successful data scientist. We see the movement toward this structure as a natural evolution of the field away from being viewed as subsets of Computer Science and Statistics/Mathematics (Figure 3 - left) to a new independent field with both cultural and intellectual uniqueness. Figure 3 (right) represents the four areas with practice being the use of these skills inside a specific area of expertise or project [1].

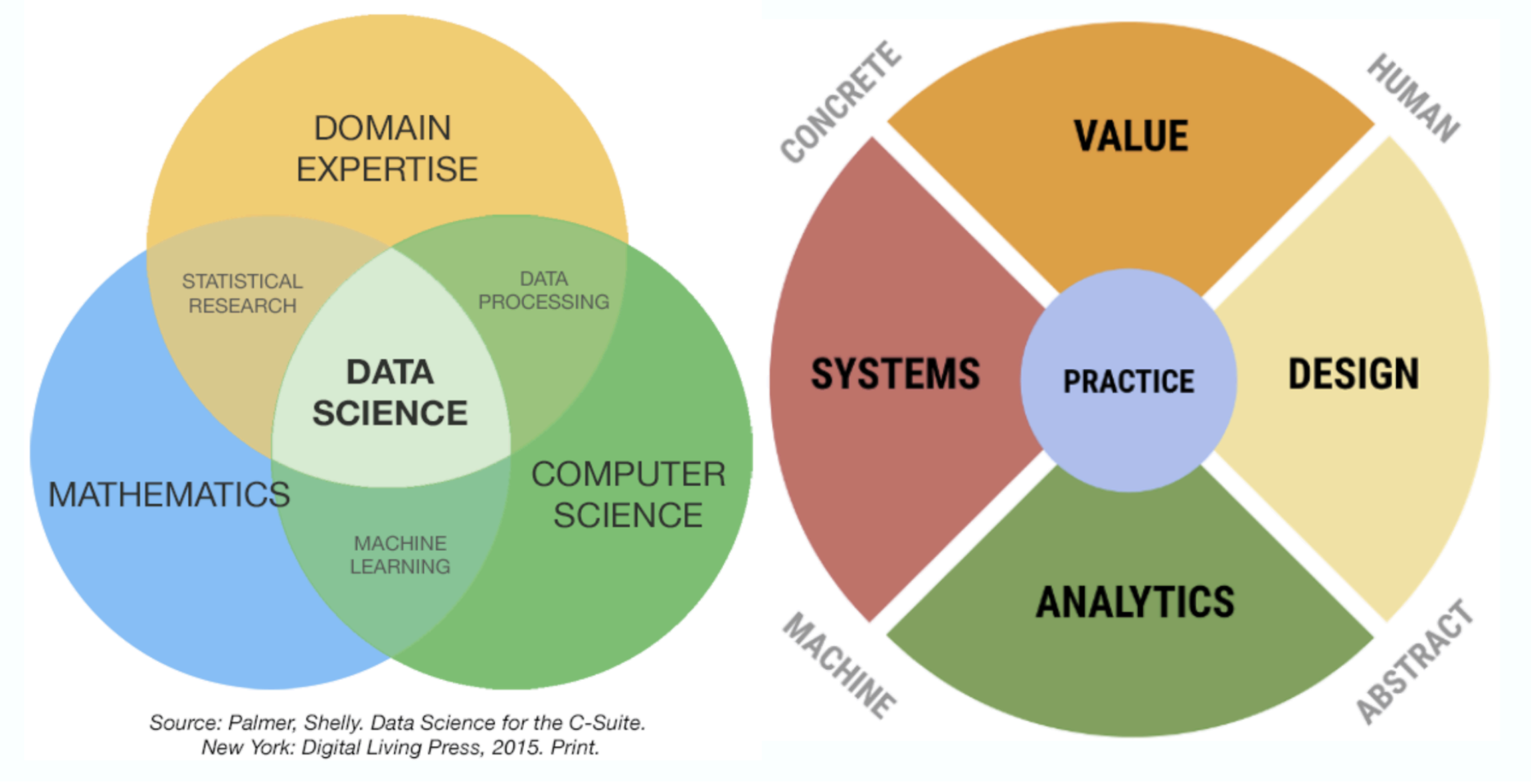

**FIGURE 3: Two Visual Representation of Data Science**

Value represents the human aspects of Data Science and inherits methods and practices from the field of ethics, but considers data oriented business practices. Cultural movements that fall into this area are widely supported across the Data Science ecosystem such as "Data for Good" or "Data and Society". Initiatives focus on "Open Data" and transparency are additional examples that would align with the overarching concept of ethical data use..

Design encompasses the creative aspects of Data Science and how we interact with data products and communicate results. Data visualization and storytelling are cornerstones along with aspects of Human Centered Design or Human Computer Interaction. The importance of this area is often understated. However, in conversations with industry and working data scientists the importance of being able to effectively communicate and show data science products is consistently mentioned as critical for success.

Systems represent the necessary tools for storing, organization, moving and processing data. Core to this area is how we represent data in databases and the tools we use to build, train and deploy algorithms. This area is closest to the field of engineering in the sense that it represents all technological infrastructure necessary for data driven products to become a reality.

Analytics represents the mathematical aspects of the field. This can come in the form of traditional statistics methods, foundation applied mathematical concepts along AI and machine learning. Correlated fields such as information theory or operations research have also contributed heavily to this area.

Practice is the connective tissue holding all the areas together. It represents the life cycle in which all the components interact to form a comprehensive data driven project. Practice also includes subject area expertise that is critical to effectively solving most, if not all, data science problems [1].

## THE BACHELOR'S OF SCIENCE CURRICULUM AT UVA (BSDS)

### Overview

On September 19th 2023 the State Council of Higher Education in Virginia (SCHEV) approved a bachelor's degree in Data Science at the University of Virginia (BSDS) [7]. The cornerstone of this degree is a novel curriculum based on the model of Data Science developed by Professor Rafael Alvarado which was outlined in the previous section. In this section we will describe the key features of this curriculum and how it brings the model of Data Science to life at the Undergraduate level.

At the University of Virginia the BSDS program resides within the School of Data Science (UVADS). During their first year the students take the two prerequisite courses for admission to the major program and apply over the summer. Those courses include DS 1001, which focuses on Foundational concepts and is unique among colleges globally, and DS 1002, which focuses on the use of computers in data science – specifically programming. The DS 1001 course is the subject of extended discussion further on in this paper.

Once admitted into the major program the students pursue a core course of study consisting of 40 credits spread across two courses each in the main areas of Data Science (Design, Value, Systems, Analytics). The core also includes a specialized mathematics series that are tailored to the needs of the program. In addition to the core courses students select one or two concentrations populated with elective courses. The concentrations come in two general forms, a further exploration of one of the areas (e.g. a concentration in analytics would go further into machine learning) or an academic area outside of the school in partnership with another department (e.g. Astronomy, Physics, or Digital Humanities) [8]. One particular course to note is DS 4022, named “Final Project”. In this course students will synthesize what they have learned over the entire curriculum. This course is also larger than just the semester in which it is taken. The students will be thinking about how they can use that time all along the way. They will then be given great agency to explore and build a data science project, which for some, will lead to future career opportunites [9].

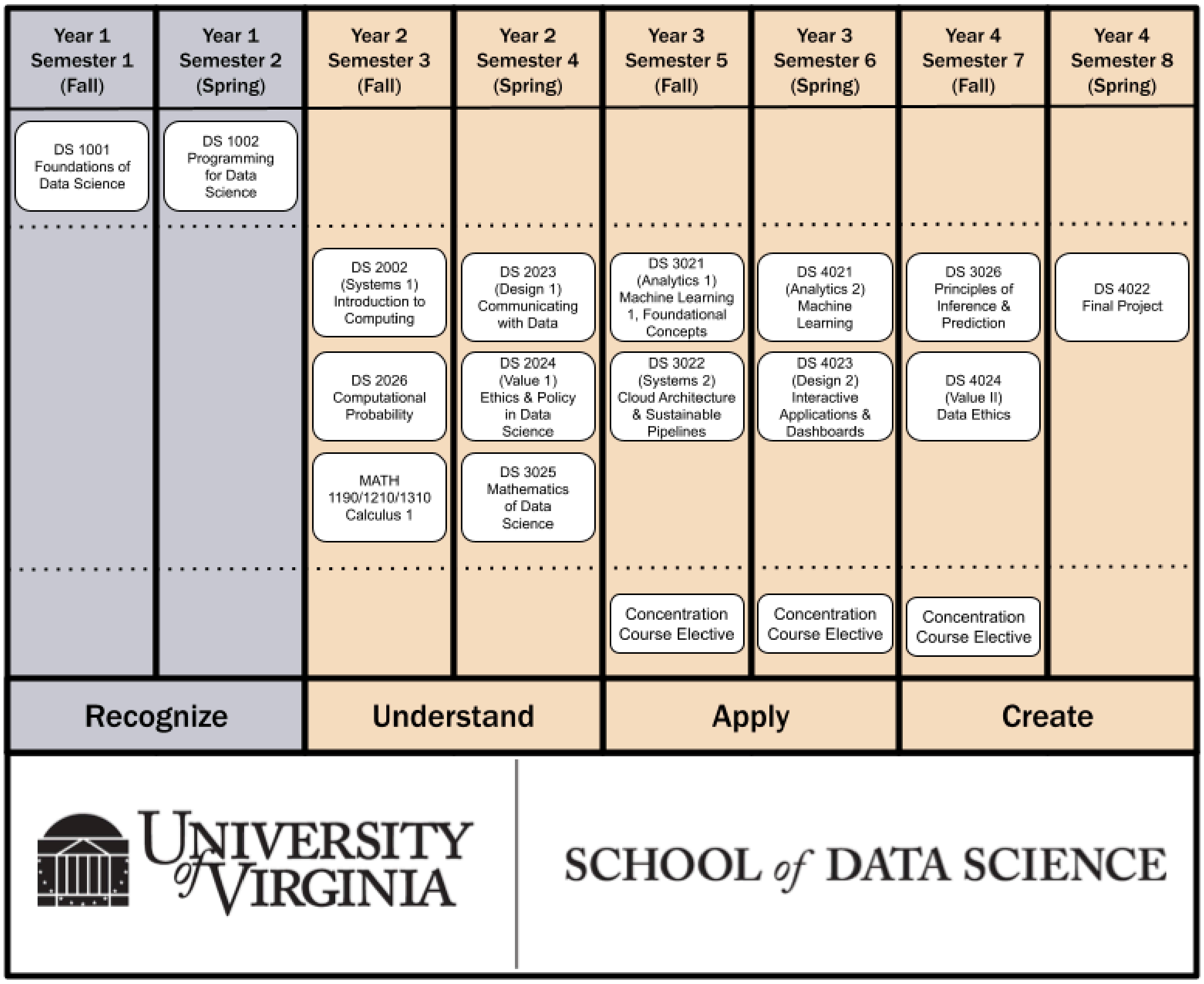

**FIGURE 4: Sample course of study for a student enrolled in the BSDS program. The first year, before the student is admitted to the School of Data Science BSDS Major Program, is indicated in blue in the first two columns, and contains the two prerequisite courses. The middle band of 13 courses covered over the 2nd, 3rd, and 4th years are the core curriculum. The final band of three courses shows the requirement for fulfilling one concentration.**

**Leadership and Teamwork**

In the rapidly evolving field of Data Science, leadership and teamwork play pivotal roles in driving innovation, fostering collaboration, and ensuring the effective utilization of data-driven insights. Data Science projects inherently involve cross-functional teams with diverse skills ranging from machine learning to domain expertise and engineering. A thoughtful leader is essential to provide a clear vision, set strategic objectives and for navigating ethical considerations and ensuring compliance with data privacy regulations.

As a result, team based learning is ubiquitous in the program with the majority of classes requiring students to work together and lead projects as core assessments. The completion of a Service Learning project is also a requirement for graduates. This can include tutoring new students, serving as a TA, or actively participating in the various community learning activities in the Charlottesville area. The common thread is that this is a new field that is growing quickly and we all need to work together to shape the present and the future of Data Science for the better.

**After Graduation**

Previously we discussed that most Data Science courses at the undergraduate level are really Computer Science or Statistics courses with a Data Science title. When we discuss the advantage that our graduate will have we focus directly on that fact. In this curriculum there are elements critical to succeeding in Data Science work that are not addressed in those other curricula. The core feature is the fundamental acknowledgement that humans are the beginning and the end of every Data Science endeavor. There is always a person who sets the goals at the start, there is always a person conducting the work along the way, and there is always a person who is impacted by the results of the project. As more and more advances are made in AI the value of the college education will shift more and more towards the graduates ability to make the deliverables fit into the human world.

In particular we highlight the lessons in leadership and teamwork. Those themes run across all human endeavors and have specific nuances which must be mastered for every specific field. Data Science is no exception and our curriculum addresses this directly in the Design and Value components.

Another critical component is the understanding of the principles of an Established Data Set. That is to say a Data Set which will stand the test of time. Concepts like provenance and licensing, as well as data management and ethical impact are explicitly taught and practiced throughout this curriculum.

The final critical differentiator for our graduates is their ability to communicate their results to other humans. Most programs do not teach data presentation and in many places it is pejoratively called a “soft-skill”. This curriculum rejects that premise and through the Design courses students are explicitly trained in the skills required to present work with a foundation in data analysis.

**DS 1001 Foundations of Data Science**

The Foundations of Data Science course sets the table for the next three years by presenting the field through UVA’s easily consumable and understood model. The end result is that students understand potential pathways for study and more importantly can envision themselves walking these pathways through the curriculum and eventually out into the world. This is accomplished through the inclusion of guest lectures, both in and out of academia, applied weekly labs and more traditional lecture sessions. The combination of these three main pedological components provides students with an understanding of the current state of the field and an initial understanding of the depth of each of the four areas. The calendar of the course is such that each area represents a fourth of the class, which is also how the core classes of the major are divided. This sends a clear message that data ethics and data communications are equally valued alongside analytics or systems. This equal balancing creates a wider spectrum of potential interests students can pursue and consequently a wider spectrum of students that will pursue the field, which is one of the main objectives of the course.

It is our hope that the structure of the class will be used as a model on other campuses to consider when creating "Introductory" type classes. The approach could also be pushed down into the K-12 school system, as the topics and concepts being presented do not assume prior knowledge of the field.  The course also uses a series of purposeful and contemporary design approaches, which are detailed below.

**Pedagogical Techniques**

In the following subsections we will describe several pedagogical techniques that are incorporated into DS 1001, indicating why they were selected, and sharing details of their implementation. The techniques we have chosen to highlight here are not an exhaustive list but rather most critical for the success of the students in the class. At the highest level each was chosen to help us achieve the prime directive of our course design, "any student can enroll and excel". The techniques detailed here are Specifications Grading, Active Learning in Lectures, Active Learning Labs with gamification, Team Teaching, and Enhanced Student Agency.

**Specifications Grading**

When constructing this course we needed to select a grading system that fit our design principles. A major component of Data Science work is that you must complete and finish your work for it to have value. If the analytics component is in place it is of no value without the systems solution to deliver it. As a result we selected a grading system that focuses on completion of projects rather than on accumulation of points. The point system, pedagogically referred to as "mean-weighted average", is most commonly used throughout American high school and college but fails to meet the completion needs [10].

The system called specifications grading is designed to foster a spirit of completing projects [11]. The stand out features are single-level rubrics coupled with assignment bundles corresponding to letter grades. In this way every student can select what letter grade they want to earn and pursue the assignments for that bundle. The single-level rubrics also create a dynamic where the vision of the task is clearly laid out for the students. They know precisely what the criteria are for their assessment. This element also ties into the student agency component discussed later.

**Active Learning Lectures**

The Foundations of Data Science course is a survey course designed for first-year college students. As a result one of the design specifications for the course is a community building objective, another is introducing students to concepts for the first time. The use of active learning activities like "think-pair-share" are well suited to addressing both of these objectives [12]. The downside of most active learning exercises is that they take more time and more engagement from the instructor. Fortunately neither of this is a problem for this course design.
Given the survey and introductory nature of the course there is no objective to "cover" as many topics as possible. Rather the goal is to land a few key questions and ideas at a deep level. This permits ample time. In addition the team teaching nature of the class, elaborated on below, helps to spread the work out and not overburden the instructors.

**Active Learning Labs with Gamification**

This course is designed with a lab component. Students enrolled in DS 1001 simultaneously enroll in a lab section. It is fully integrated with the lecture portion of the course and labs are delivered in sequence with the material in the lecture portion of the course. In these lab sections students are given time to explore practical examples of the topics discussed in lecture. A key feature of these labs is the use of gamification.

Many children's games, such as Battleship or Mastermind, are elementary puzzles based directly on principles taught in the lecture portion of the class. When the students play the games with each other in the lab period they naturally ask questions about how the game works and what strategies will be the most sucessfull. The students are able to explore the topics from class in a self-directed way and find solutions to the puzzles presented by the games by tapping into the knowledge they gained in class [13] [14]. (As a side-note: some distinguish between "gamification" and "game based learning", for our purposes here we use them interchangeably).

**Team Teaching**

A fundamental principle of Data Science is that it is a "Team Sport". As a result the design of this course was mandated to demonstrate that principle wherever possible. To that end the design and delivery of the course was conducted by a team of instructors in discussion with the wider faculty of the school. During a typical semester the students will interact with: two primary co-instructors, a staff data scientist, the TA for their lab section, four additional professors invited to give guest lectures, and four industry experts invited to give guest lectures.
There are several benefits from the team approach. The first is modeling to the students the principles of teamwork in data science. The second is to offer them diverse views and demonstrate that there are many ways to look at an issue. On the backend side the division of labor allows for specialization and reduces the workload on any individual instructor. This allows the instructors to focus on developing individual assignments at a much higher quality than normal. The downside of this approach is the increased time of coordination. This is a real concern and requires the team to function well together.

**Enhanced Student Agency**

The final design goal that we want to highlight here is the goal of empowering students. The primary mechanism we use is by giving the student agency in the classroom when possible, and at an appropriate level. The most explicit way this is done is by giving the students the choice of which grade they set as their goal. The requirements for each letter grade are mapped onto a specific bundle of assignments. Traditional classrooms mandate completion of every assignment and produce an aggregate score at the end.

Another key aspect of student agency are the deep dive assignments. The “look ahead” and “case study extension” assignments are designed to be deep dives where a student can fully explore an idea. Over the course four of each type are offered for a total of eight assignments. However the “A” bundle only requires four to be completed. The students can take their own career goals in mind when choosing which assignments to complete. And is this way able to tailor their learning to meet their needs.

## CONCLUSION

The field of Data Science has rapidly appeared on university campuses nationwide and is a remarkably young academic field, by some measures only a decade or so old. The next iteration will be the creation of undergraduate programs at likely hundreds of schools across the country, if not the globe, over the next three to five years. The curriculum content of those programs could shape the field for generations, making it imperative that as a community we work together to develop a consensus on what should be first principles for the field. It is in this context that we worked to create the overarching framework (4+1 model), an associated Bachelor of Science curriculum, and a broad introductory course whose goal is to present more accurately the field of Data Science as an independent discipline with unique knowledge and skill requirements. Our hope is that this will trigger a larger conversation, or at a minimum encourage some reflection for those confronted with building or shaping Data Science academic programs.